\documentclass[letter]{aa}
\usepackage{times}
\usepackage{natbib}
\usepackage{graphicx}
\usepackage{supertabular}

\begin{document}
\title{Determination of the light curve of the Rosetta target Asteroid (2867) Steins
by the OSIRIS cameras onboard Rosetta}
\author{M. K\"uppers\inst{1} \and S. Mottola\inst{2} \and S. C. Lowry\inst{3} \and M. F. A'Hearn\inst{4}  \and C. Barbieri\inst{5} \and M. A. Barucci\inst{6} \and
        S. Fornasier \inst{5,6} \and O. Groussin\inst{4,7} \and P. Guti\'errez \inst{8} \and S. F. Hviid
        \inst{1} \and H. U. Keller \inst{1} \and P. Lamy \inst{7}}
\institute{Max-Planck-Institut f\"ur Sonnensystemforschung, Germany, kueppers@mps.mpg.de \and
           Institute of Planetary Research, DLR, Germany \and Astrophysics Research Centre, Queen's University Belfast,
           United Kingdom \and Department of Astronomy, University of Maryland, USA \and Dipartimento di Astronomia and CISAS, Universit\`a di Padova, Italy
           \and University of Paris 7, France
           \and Laboratoire d'Astrophysique de Marseille, France \and Instituto de Astrof\'isica de Andaluc\'ia - CSIC, Spain}
\date{Received   /   Accepted}
\abstract
{In 2004 asteroid (2867) Steins has been selected as a flyby target for the Rosetta mission. Determination
of its spin period and the orientation of its rotation axis are essential for optimization of the flyby
planning.}
{Measurement of the rotation period and light curve of asteroid (2867) Steins at a phase angle larger than achievable
from ground based observations, providing a high quality data set
to contribute to the determination of the orientation of the spin axis and of the pole direction.}
{On March 11, 2006, asteroid (2867) Steins was observed
continuously for 24 hours with the scientific camera system OSIRIS onboard Rosetta. The phase angle was
41.7
degrees, larger than the maximum phase angle of 30 degrees when Steins is observed from Earth. A total of
238 images, covering four rotation periods without interruption,  were acquired.}
{The light curve of (2867) Steins is double peaked with
an amplitude of $\approx$ 0.23 mag. The rotation period is 6.052 $\pm$ 0.007 hours. The continuous observations
over four rotation periods exclude the possibility of period ambiguities. There is no indication of deviation
from a principal axis rotation state. Assuming a slope parameter of  G = 0.15, the absolute visual magnitude of
Steins is 13.05 $\pm$ 0.03.   }
{}
\keywords{minor planets, asteroids -- techniques: photometric}

\titlerunning{OSIRIS observations of Asteroid (2867) Steins}
\authorrunning{K\"uppers et al.}

\maketitle

\section{Introduction}
Rosetta is a Cornerstone mission of the European Space Agency (ESA).
Its goal is to monitor a comet over several months on its way to perihelion
with an orbiter and to investigate its surface and interior with a lander.
Additionally, flybys of 2 asteroids are part of the mission plan.

The launch of Rosetta, first planned for January
2003, had to be postponed to the first quarter of 2004 due to problems with
the launcher. As a consequence, all 3 mission targets had to be changed.
Comet 67P/Churyumov-Gerasimenko was chosen as the new target comet.
After the successful launch on 2 March 2004 the asteroids (2867) Steins and
(21) Lutetia were selected as flyby targets, with closest approach in Sept.
2008 and July 2010, respectively.

While the properties of the large main belt asteroid (21) Lutetia are
relatively well known, (2867) Steins had not been investigated in detail
before it was selected as a spacecraft target. Since knowledge of the
global properties of the target is essential for the success of the flyby,
several observing programs of Steins have been initiated starting in
2004.

Steins has been classified as an E-type asteroid based on its visual and
near-infrared spectrum \citep{Barucci05} and on its high albedo, as determined
from its polarimetric properties \citep{Fornasier06}. The polarimetric and spectral
 properties of Steins imply an extensive
thermal history of a differentiated body. Steins may have a composition
similar to relatively rare enstatite chondrite/achondrite meteorites. With an absolute
magnitude of 13.18 \citep{Hicks04}, the polarimetric
albedo of 0.45 $\pm$ 0.10 corresponds to a diameter of approximately 4.6 km.

\citet{Hicks04} reported a rotation period of 6.06 $\pm$ 0.05 hours for
Steins, which was followed up with a full detailed analysis by \citet{Weissman05},
resulting in a more precise value of 6.048 $\pm$ 0.007 hours for the synodic period. The
light curve amplitude on both occasions was 0.20-0.29 magnitudes. The colours
determined by \citet{Hicks04} were V-R = 0.58 $\pm$ 0.03, R-I = 0.44 $\pm$ 0.03,
and B-V = 0.80 $\pm$ 0.03, while \citet{Weissman05} derived V-R = 0.51 $\pm$ 0.03 and
R-I = 0.44 $\pm$ 0.03.

Here we report on observations of Steins obtained with the scientific optical camera
system OSIRIS onboard Rosetta. The data set taken from a spacecraft in interplanetary
space  is unique in terms of its continuous coverage over 24 hours and in the
phase angle of 41.7 degrees, which is larger than the maximum phase angle of
Steins that can be seen from Earth (30 degrees).

\section{Instrument, Observations and Data reduction}
OSIRIS consists of a narrow angle camera (NAC) and a wide angle camera (WAC).
They are unobstructed mirror systems with focal lengths of 72 cm (NAC) and
14 cm (WAC). The focal ratio is  f/D = 8 and f/D = 5.6 for the NAC and the
WAC, respectively. Both cameras are equipped with 2048 x 2048 pixel CCD
detectors with a pixel size of 13.5$\mu$m. The image scale is 3.9 arcsec/pixel
for the NAC and 20.5 arcsec/pixel for the WAC. OSIRIS is described in detail in
\citet{Keller06}.

\begin{figure}
\resizebox{\hsize}{!}{\includegraphics[width=\textwidth]{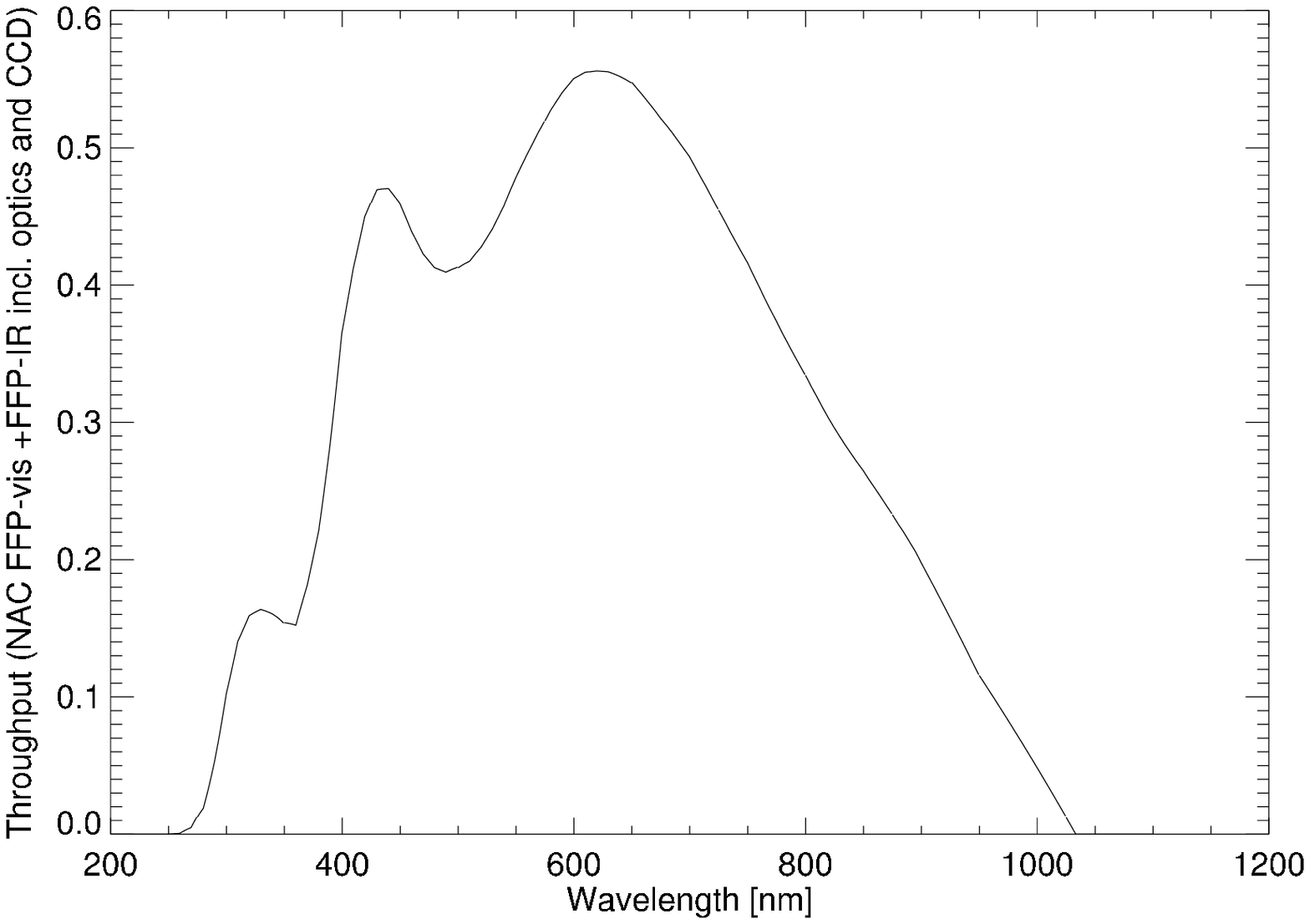}}
\caption{System throughput of the OSIRIS NAC with clear filter, including
the transmission of the mirrors and of the visual and near-infrared re-focussing
plate, and the quantum efficiency of the CCD. The cut-off at near-ultraviolet
wavelengths is due to the near-IR focus plate becoming opaque, the cut-off in
the near-IR is caused by the decreasing quantum efficiency of the CCD.}
\label{through}
\end{figure}

The NAC was used for the observations of (2867) Steins because of its higher
point source sensitivity compared to the WAC. The NAC is equipped with two
filter wheels, containing 8 positions each. This allows for numerous
combinations of 11 medium bandwidth filters, 4 re-focussing plates, and a
neutral density filter. For the observations of (2867) Steins the re-focussing
plates for the visual and for the near-infrared were used, effectively
providing a clear filter. The system throughput in this configuration is
shown in Fig. \ref{through}.

The observations took place on 11 March 2006. The distance of (2867) Steins
from Rosetta and from the Sun was 1.06 AU and 2.30 AU, respectively. The
phase angle of Steins was 41.7 degrees. The Rosetta spacecraft tracked the
expected motion of Steins. A total of 238 images with 300 s exposure time
each were obtained, one image every 6 minutes. A window of 512 x 256 CCD
pixels centered on the expected position of Steins was read out.

The data were reduced with the OSIRIS standard calibration pipeline. For
the Steins observations, the pipeline performed the following steps:
\begin{itemize}
\item The CCD is read out using a dual 14 bit Analogue to Digital Converter
(ADC). DN values created by the ``upper'' ADC (DN values $\ge$ 2$^{14}$) get
an additional offset added by the read out electronics. This offset is
subtracted.
\item A noise filter removes coherent noise in the data produced by the
power converter.
\item The bias is subtracted using bias exposures taken immediately after
      the Steins observations.
\item The images are divided by a flat field using reference flat fields from
      the ground calibration.
\item The data are converted into units of W m$^{-2}$nm$^{-1}$sterad$^{-1}$, using
      a conversion factor derived from observations of Vega, the secondary
      spectrophotometric standard stars 58 Aql and \mbox{$\epsilon$ Aqr,} and of the
      solar analog stars 16 Cyg A and 16 Cyg B. This conversion is valid for a solar
      source spectrum.
\end{itemize}

\begin{figure}
\resizebox{\hsize}{!}{\includegraphics{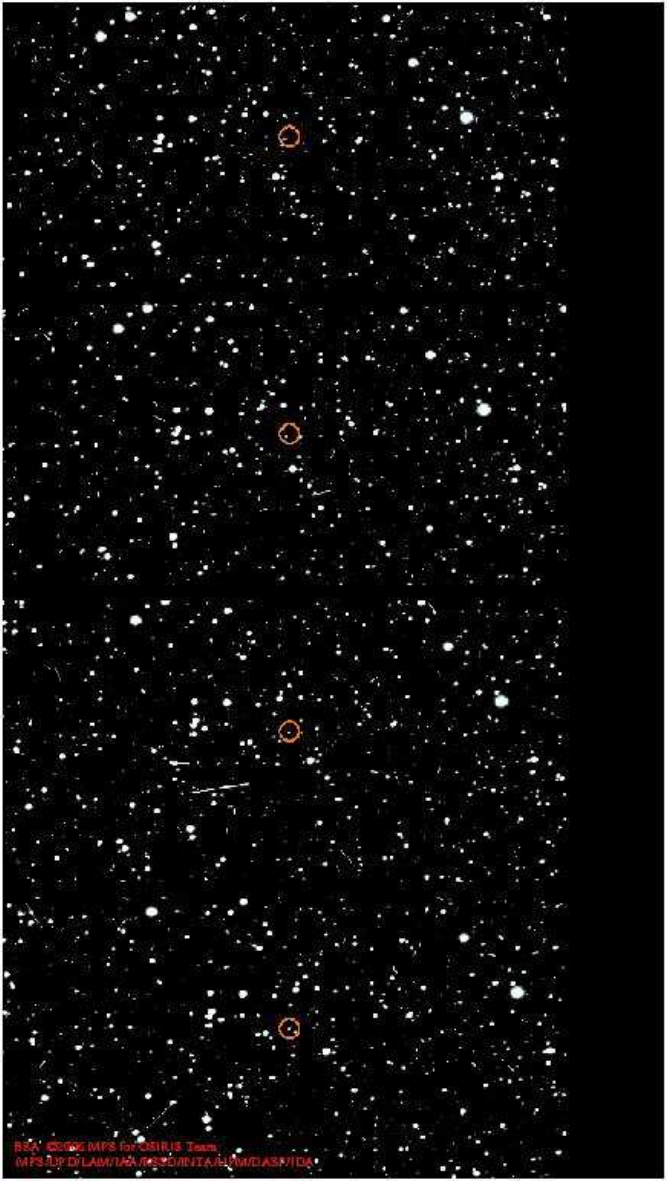}}
\caption{Four images of asteroid (2867) Steins. The images in the second, third, and
fourth panel, from bottom to top, were taken approximately 5, 10, and 15 hours after the first
image, respectively. In all images Steins can be seen within the orange circle, demonstrating
the excellent tracking of the spacecraft.}
\label{tracking}
\end{figure}

\begin{figure}
\resizebox{\hsize}{!}{\includegraphics{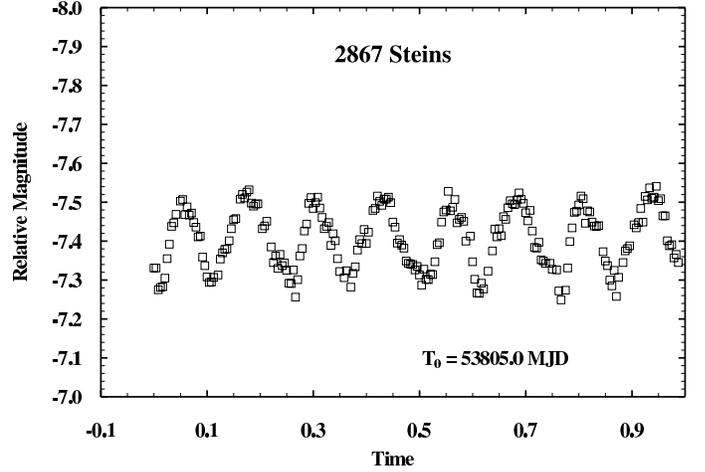}}
\caption{All OSIRIS observations of asteroid Steins, covering four rotation periods.}
\label{rawmag}
\end{figure}

After the systematic processing pipeline had been applied,
we carried out the photometric reduction procedure, aimed at
extracting the flux information from the images.
During the observation period Steins moved across the NAC field of view,
and it appulsed background stars in many instances.
For this reason, the first step of the photometric reduction consisted of
modeling the star background and its subsequent subtraction.
To this end, all of the science images were resampled, registered
and co-added in five groups, to provide an accurate background
representation for different segments of the asteroid's trajectory.
During this process, transient spurious events (``cosmic ray hits'') have been
discriminated by comparison and removed. Subsequently, the corresponding
background fields have been subtracted from each science frame. The subtraction
in image space (as opposed to subtraction of total flux) has been made possible by
the fact that the spacecraft tracking was very accurate (see Fig. \ref{tracking}),
with an error within a single exposure smaller than the PSF of the camera. Finally, the asteroid flux
was extracted from the background-subtracted images by means of a synthetic
aperture procedure in which the pixel intensities are integrated over
a user-defined aperture. The phase angle varied between 41.59\degr and 41.83\degr during
the observations. Since the brightness change induced by the phase angle change is
expected to be less than 0.01 mag, no phase correction has been performed.
The resulting light curve is shown in Table \ref{magtime} and Figure \ref{rawmag}.

\section{Results and Discussion}
\begin{figure}
\includegraphics[width=\linewidth]{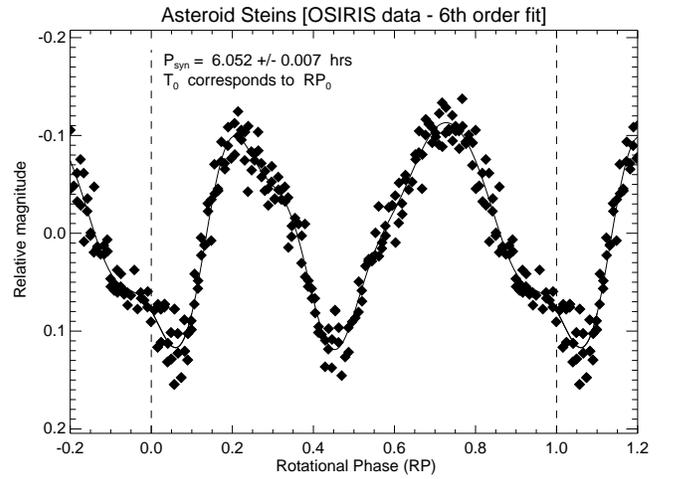}
\caption{The light curve of asteroid (2867) Steins from OSIRIS data. Magnitudes are
given with arbitrary zero point. $T_0$ is \mbox{53804.99981 MJD}, the data at phases
less than zero and larger than one are repeated for clarity. }
\label{lightcurve}
\end{figure}

The resulting photometric time series have been converted to magnitudes and
modelled as a 6th order Fourier polynomial \citep{Harris89a}, whose best fit gave a
synodic rotation period of \mbox{6.052 $\pm$ 0.007 hours}. The
data folded with the best fit period are  shown in Fig.\ref{lightcurve}.

The figure demonstrates that the light curve of (2867) Steins is double peaked, with an amplitude of $\approx$ 0.23 mag.
The rotation period is 6.052 $\pm$ 0.007 hours, which fits previous determinations well \citep{Hicks04,Weissman05}.
The continuous observations over four rotation periods exclude the possibility of period ambiguities. The behaviour of the light curve
is not symmetric and the differences between  the two maxima and two minima would imply an irregular shape for
the asteroid. The shoulder on the descending portion of the light curve near the second minimum is reminiscent on a similar feature
in the light curve of comet 9P/Tempel 1 \citep{A'Hearn05}. In the case of Tempel 1 the shoulder was caused by large, relatively flat
surfaces abutting each other.

There is no indication of a deviation from a principal axis rotation state. Therefore, the OSIRIS data,
taken together with ground based data obtained at different epochs, can be used to derive a solution
for pole orientation \citep{Lamy06}.

If the rotational brightness variation were caused by a smooth, featureless
triaxial ellipsoid that scatters light geometrically, then the diameter ratio $a/b$ between the
longest and the intermediate semiaxis is $>$ 1.23.

The average clear filter brightness of asteroid (2867) Steins was converted into absolute brightness.
The image calibration provides the flux from Steins in W m$^{-2}$ nm$^{-1}$ at a wavelength of 600 nm,
assuming a solar spectrum. First the flux was corrected for the $\approx$ 12.3 \% of the light
that is outside the photometric aperture with two pixels radius due to the point spread function of the
camera. Then the total flux was corrected for the spectrum of
Steins being redder than that of the Sun using
\begin{equation}
F_c = F_u \frac{\int_\lambda F_{\sun} (\lambda) T(\lambda)\,d\lambda}{\int_\lambda F_{\rm{steins}} (\lambda) T(\lambda)\,d\lambda}
\end{equation}
Here $F_c$ and $F_u$ are the corrected and uncorrected flux of asteroid
Steins at 600 nm, $T(\lambda)$ is the system throughput shown in Fig.
\ref{through}, and $F_{\sun} (\lambda)$ and $F_{\rm{steins}} (\lambda)$ are
the spectra of the Sun and of Steins, respectively, both normalized to unity
at 600 nm. The solar spectrum was taken from \citet{Burlov-Vasiljev95} and
\citet{Burlov-Vasiljev98} in the visible range, and from \citet{Woods96} in the
ultraviolet range ($<$ 350 nm). The disk-center spectra from Burlov-Vasiljev
et al. were corrected for limb darkening using the co-efficients of
\citet{Neckel94}. The spectrum of Steins was taken from \citet{Barucci05}. We
assumed a constant albedo of Steins below 390\,nm where, to our knowledge, no
spectrum exists. Anyhow, the contribution of the ultraviolet spectral range
to the flux measured through the clear filter is low.

$F_c$ = (
8.70 $\pm$ 0.15) $\times$ 10$^{-18}$ W m$^{-2}$ nm$^{-1}$ is about 2.5 \% higher than
$F_u$. With $F_c$ and the spectrum of \citet{Barucci05} we know the absolute spectral
flux distribution of Steins. Then the visual magnitude of Steins is given by:
\begin{equation}\label{magnitude}
m_V = -2.5 \log(F_c \int_\lambda F_{\rm{steins}} (\lambda)
T_V(\lambda)\,d\lambda)+ C
\end{equation}
Here $T_V$ is the transmission of the standard V filter taken from
\citet{Bessel90}. We determine the constant $C$ as the difference between
the apparent magnitude of the Sun of -26.75 \citep{Cox2000}
and the convolution of the solar spectrum with the V filter as in eq. \ref{magnitude}.
The apparent magnitude of asteroid (2867) Steins is V(r,$\Delta$,$\alpha$) =
 16.61 $\pm$ 0.03. The corresponding magnitude
at heliocentric and observer distance of 1 AU would be V(1,1,$\alpha$) = 14.68. The conversion to
absolute magnitude depends on the phase function of the asteroid,
parameterized with the slope parameter $G$ \citep{Bowell89}. For the standard assumption of $G$ =
0.15, the absolute magnitude is 13.05 $\pm$ 0.03.  With an albedo between 0.3 and 0.45
\citep{Fornasier06,Lamy06}, the absolute magnitude of  13.05
corresponds to an effective diameter between 4.8\,km and 6.0\,km.

\section{Summary}
Table \ref{Results} summarizes the results of the present paper and compares
them to the photometry of \citet{Hicks04} and \citet{Weissman05}.
The agreement on the rotation period is excellent.

The absolute brightness of (2867) Steins determined from the OSIRIS
measurements is $\approx$ 13 \% higher than that derived from
ground based measurements. The reason could be the photometric slope
parameter $G$, which in E type asteroids is normally larger than the
standard value of 0.15 ($\approx 0.5$, e.g. \citet{Harris89b}). Since the
phase angle of (2867) Steins during the OSIRIS observations was larger than
during the observations of \citet{Hicks04} (11.5\degr) and \citet{Weissman05}
(17 \degr), the data would be most consistent with $G \approx 0.3$ and an
absolute brightness of $\approx$ 13.3. Additionally, the absolute brightness
of an asteroid with a light curve of moderate amplitude varies with
aspect angle by a few tenth of a magnitude \citep{Karttunen89}. Since the
position of Steins on the sky differed by more than 90 degrees between the
ground based observations from 2004 \citep{Hicks04,Weissman05} and the
OSIRIS observations in 2006, it is likely that the aspect angle was
indeed different.

\begin{table}
\caption{Comparison of photometry of asteroid (2867) Steins. $H$ is the
         absolute magnitude with an assumed slope parameter of 0.15.}
\label{Results}
\centering
\begin{minipage}{\textwidth}
\renewcommand{\footnoterule}{\rule{0cm}{0cm}}
\begin{tabular}{c c c c }
\hline\hline
& This work & H04\footnote{\citet{Hicks04}} & W05\footnote{\citet{Weissman05}}\\
\hline
Period [hrs] & 6.052 $\pm$ 0.007  & 6.06 $\pm$
0.05 & 6.048 $\pm$ 0.007\\
Amplitude [mag] & 0.23 $\pm$ 0.03 & 0.2 & 0.29 $\pm$ 0.04 \footnote{in R filter} \\
H [mag] & 13.05 $\pm$ 0.03 & 13.18 $\pm$ 0.04 & 13.18 $\pm$ 0.02 \\
\hline
\end{tabular}
\end{minipage}
\end{table}

The nominal flyby with the asteroid Steins will take place on September
5th, 2008 at a velocity of 8.565 km/s and a close approach of 1745 km. The
obtained results from OSIRIS will be fundamental together with ground
based observational campaigns  to define the shape and pole orientation of the
object. These properties will allow to optimize the spacecraft geometry
during the Steins flyby to maximize the scientific return of
the Rosetta mission.

Finally, these results attest to the excellent capabilites of the OSIRIS camera
in terms both of scientific usefulness and technical performance.

\begin{acknowledgements}
The OSIRIS imaging system on board Rosetta is managed by the
Max-Planck-Intitute for Solar System Research in Katlenburg-Lindau (Germany),
thanks to an International collaboration between Germany, France, Italy,
Spain, and Sweden. The support of the national funding agencies DLR, CNES,
ASI, MEC, and SNSB is gratefully acknowledged. We acknowledge the work of the Rosetta Science Operations
Centre at ESA/ESTEC and of the Rosetta Mission Operations Centre at ESA/ESOC
who made these observations possible on short notation and
operated the spacecraft. S.C.L. acknowledges support from the Leverhulme Trust.
This research made use of JPL's online ephemeris generator (HORIZONS).

\end{acknowledgements}

\Online

\twocolumn
\centering
\tablecaption{\label{magtime} Flux from Asteroid Steins measured by OSIRIS. MJD is the modified Julian Date corrected
for light-travel time, mag is
the magnitude measured in the clear filter with arbitrary zero point. The letter X in the first
column denotes measurements that are uncertain due to contamination by stars.}
\tablefirsthead{\hline\hline
& MJD & Mag \\
\hline}
\tablehead{\multicolumn{3}{c}{{\bfseries Table \ref{magtime}} (continued)} \\
\hline\hline
& MJD & Mag \\
\hline}
\begin{supertabular}{ c c c}
          &   53804.99981       &  -7.331\\
          &   53805.00395       &  -7.331\\
          &   53805.00812       &  -7.275\\
          &   53805.01228       &  -7.281\\
          &   53805.01645       &  -7.283\\
          &   53805.02062       &  -7.305\\
          &   53805.02478       &  -7.355\\
          &   53805.02895       &  -7.392\\
          &   53805.03312       &  -7.438\\
          &   53805.03728       &  -7.447\\
          &   53805.04145       &  -7.469\\
X         &   53805.04562       &  -7.539\\
          &   53805.04978       &  -7.503\\
          &   53805.05395       &  -7.507\\
          &   53805.05812       &  -7.468\\
          &   53805.06228       &  -7.488\\
          &   53805.06645       &  -7.467\\
          &   53805.07062       &  -7.472\\
          &   53805.07479       &  -7.448\\
          &   53805.07896       &  -7.436\\
          &   53805.08313       &  -7.411\\
          &   53805.08729       &  -7.413\\
          &   53805.09146       &  -7.359\\
          &   53805.09563       &  -7.337\\
          &   53805.09979       &  -7.308\\
          &   53805.10396       &  -7.294\\
          &   53805.10813       &  -7.295\\
          &   53805.11229       &  -7.307\\
X         &   53805.11646       &  -7.272\\
          &   53805.12063       &  -7.313\\
          &   53805.12479       &  -7.354\\
          &   53805.12896       &  -7.370\\
          &   53805.13313       &  -7.379\\
          &   53805.13729       &  -7.380\\
          &   53805.14146       &  -7.401\\
          &   53805.14563       &  -7.432\\
          &   53805.14979       &  -7.455\\
          &   53805.15396       &  -7.458\\
X         &   53805.15813       &  -7.420\\
          &   53805.16229       &  -7.508\\
          &   53805.16646       &  -7.520\\
          &   53805.17063       &  -7.511\\
          &   53805.17479       &  -7.526\\
          &   53805.17896       &  -7.532\\
          &   53805.18313       &  -7.498\\
          &   53805.18729       &  -7.490\\
          &   53805.19146       &  -7.495\\
          &   53805.19563       &  -7.496\\
X         &   53805.19979       &  -7.567\\
          &   53805.20396       &  -7.432\\
          &   53805.20813       &  -7.439\\
          &   53805.21229       &  -7.451\\
X         &   53805.21646       &  -7.328\\
          &   53805.22064       &  -7.385\\
          &   53805.22479       &  -7.345\\
          &   53805.22896       &  -7.362\\
          &   53805.23313       &  -7.330\\
          &   53805.23729       &  -7.366\\
          &   53805.24146       &  -7.337\\
          &   53805.24563       &  -7.344\\
          &   53805.24979       &  -7.325\\
          &   53805.25396       &  -7.292\\
          &   53805.25813       &  -7.291\\
          &   53805.26229       &  -7.326\\
          &   53805.26646       &  -7.256\\
          &   53805.27063       &  -7.301\\
          &   53805.27479       &  -7.362\\
          &   53805.27896       &  -7.381\\
          &   53805.28313       &  -7.426\\
          &   53805.28729       &  -7.444\\
          &   53805.29146       &  -7.497\\
          &   53805.29563       &  -7.512\\
          &   53805.29979       &  -7.484\\
          &   53805.30396       &  -7.499\\
          &   53805.30813       &  -7.513\\
          &   53805.31229       &  -7.485\\
          &   53805.31646       &  -7.461\\
          &   53805.32063       &  -7.432\\
          &   53805.32479       &  -7.438\\
          &   53805.32896       &  -7.448\\
          &   53805.33313       &  -7.389\\
          &   53805.33729       &  -7.419\\
          &   53805.34146       &  -7.401\\
          &   53805.34563       &  -7.355\\
          &   53805.34979       &  -7.322\\
X         &   53805.35396       &  -7.501\\
          &   53805.35813       &  -7.306\\
          &   53805.36229       &  -7.324\\
X         &   53805.36646       &  -7.688\\
          &   53805.37063       &  -7.282\\
          &   53805.37479       &  -7.317\\
          &   53805.37896       &  -7.334\\
          &   53805.38313       &  -7.377\\
          &   53805.38731       &  -7.404\\
          &   53805.39146       &  -7.394\\
          &   53805.39563       &  -7.430\\
          &   53805.39979       &  -7.394\\
          &   53805.40396       &  -7.423\\
X         &   53805.40813       &  -7.549\\
          &   53805.41229       &  -7.479\\
          &   53805.41646       &  -7.484\\
          &   53805.42063       &  -7.516\\
          &   53805.42479       &  -7.502\\
          &   53805.42896       &  -7.492\\
          &   53805.43313       &  -7.508\\
          &   53805.43729       &  -7.508\\
          &   53805.44146       &  -7.513\\
          &   53805.44563       &  -7.499\\
          &   53805.44979       &  -7.449\\
          &   53805.45396       &  -7.436\\
          &   53805.45813       &  -7.395\\
          &   53805.46229       &  -7.404\\
          &   53805.46646       &  -7.390\\
          &   53805.47063       &  -7.382\\
          &   53805.47479       &  -7.349\\
          &   53805.47896       &  -7.343\\
          &   53805.48313       &  -7.341\\
          &   53805.48729       &  -7.340\\
          &   53805.49146       &  -7.326\\
          &   53805.49563       &  -7.335\\
          &   53805.49979       &  -7.313\\
          &   53805.50396       &  -7.287\\
          &   53805.50813       &  -7.328\\
          &   53805.51229       &  -7.302\\
          &   53805.51646       &  -7.301\\
          &   53805.52063       &  -7.315\\
          &   53805.52479       &  -7.314\\
          &   53805.52896       &  -7.347\\
          &   53805.53313       &  -7.391\\
          &   53805.53729       &  -7.396\\
          &   53805.54146       &  -7.449\\
          &   53805.54563       &  -7.475\\
          &   53805.54979       &  -7.480\\
          &   53805.55396       &  -7.528\\
          &   53805.55813       &  -7.478\\
          &   53805.56229       &  -7.487\\
          &   53805.56646       &  -7.507\\
          &   53805.57063       &  -7.447\\
          &   53805.57479       &  -7.456\\
          &   53805.57896       &  -7.461\\
          &   53805.58313       &  -7.451\\
          &   53805.58729       &  -7.400\\
X         &   53805.59146       &  -7.512\\
          &   53805.59563       &  -7.413\\
          &   53805.59979       &  -7.347\\
          &   53805.60396       &  -7.302\\
          &   53805.60813       &  -7.267\\
          &   53805.61229       &  -7.266\\
          &   53805.61646       &  -7.292\\
          &   53805.62063       &  -7.277\\
X         &   53805.62479       &  -7.504\\
          &   53805.62896       &  -7.323\\
X         &   53805.63313       &  -7.563\\
          &   53805.63729       &  -7.375\\
          &   53805.64146       &  -7.431\\
          &   53805.64563       &  -7.411\\
          &   53805.64979       &  -7.432\\
          &   53805.65396       &  -7.414\\
          &   53805.65813       &  -7.463\\
          &   53805.66229       &  -7.456\\
          &   53805.66646       &  -7.486\\
          &   53805.67063       &  -7.504\\
          &   53805.67479       &  -7.496\\
          &   53805.67896       &  -7.494\\
          &   53805.68313       &  -7.506\\
          &   53805.68729       &  -7.524\\
          &   53805.69146       &  -7.508\\
          &   53805.69563       &  -7.497\\
          &   53805.69979       &  -7.473\\
          &   53805.70396       &  -7.452\\
          &   53805.70813       &  -7.479\\
          &   53805.71229       &  -7.426\\
          &   53805.71647       &  -7.384\\
          &   53805.72064       &  -7.382\\
          &   53805.72479       &  -7.397\\
          &   53805.72896       &  -7.352\\
          &   53805.73313       &  -7.349\\
          &   53805.73729       &  -7.343\\
X         &   53805.74146       &  -7.487\\
          &   53805.74563       &  -7.343\\
          &   53805.74979       &  -7.328\\
X         &   53805.75396       &  -7.550\\
          &   53805.75813       &  -7.326\\
          &   53805.76229       &  -7.272\\
          &   53805.76646       &  -7.249\\
X         &   53805.77063       &  -7.521\\
          &   53805.77479       &  -7.274\\
          &   53805.77896       &  -7.331\\
          &   53805.78313       &  -7.399\\
          &   53805.78729       &  -7.434\\
          &   53805.79146       &  -7.474\\
          &   53805.79563       &  -7.477\\
          &   53805.79979       &  -7.493\\
          &   53805.80396       &  -7.516\\
          &   53805.80813       &  -7.509\\
          &   53805.81229       &  -7.446\\
          &   53805.81646       &  -7.478\\
          &   53805.82063       &  -7.475\\
          &   53805.82479       &  -7.449\\
          &   53805.82896       &  -7.439\\
          &   53805.83313       &  -7.438\\
          &   53805.83729       &  -7.440\\
X         &   53805.84146       &  -7.549\\
          &   53805.84563       &  -7.373\\
          &   53805.84979       &  -7.349\\
          &   53805.85396       &  -7.337\\
          &   53805.85813       &  -7.300\\
          &   53805.86229       &  -7.285\\
          &   53805.86646       &  -7.325\\
          &   53805.87063       &  -7.258\\
          &   53805.87479       &  -7.307\\
X         &   53805.87896       &  -7.603\\
          &   53805.88313       &  -7.345\\
          &   53805.88729       &  -7.375\\
          &   53805.89146       &  -7.381\\
          &   53805.89564       &  -7.388\\
          &   53805.90396       &  -7.442\\
          &   53805.90813       &  -7.434\\
          &   53805.91229       &  -7.448\\
          &   53805.91646       &  -7.484\\
          &   53805.92063       &  -7.449\\
          &   53805.92479       &  -7.515\\
          &   53805.92896       &  -7.506\\
          &   53805.93313       &  -7.537\\
          &   53805.93729       &  -7.510\\
          &   53805.94146       &  -7.513\\
          &   53805.94563       &  -7.541\\
          &   53805.94979       &  -7.504\\
          &   53805.95396       &  -7.509\\
          &   53805.95813       &  -7.465\\
          &   53805.96229       &  -7.465\\
          &   53805.96646       &  -7.401\\
          &   53805.97063       &  -7.388\\
          &   53805.97479       &  -7.391\\
          &   53805.97896       &  -7.357\\
          &   53805.98313       &  -7.366\\
          &   53805.98729       &  -7.345\\

\hline

\end{supertabular}

\end{document}